# Active Learning of a Neural Network Potential for Large-Scale Atomistic Simulations of Polymer Electrolyte Membranes

*Yuta Yoshimoto,* Naoki Matsumura, Meguru Yamazaki, Yuto Iwasaki, Hiroshi Nakao, and Yasufumi Sakai*

Fujitsu Research, Fujitsu Limited, 4-1-1 Kamikodanaka, Nakahara-ku, Kawasaki, Kanagawa 211-8588, Japan

## ABSTRACT

Machine learning interatomic potentials (MLIPs) can achieve near density-functional-theory (DFT) accuracy at force-field computational cost; however, long-time, large-scale molecular dynamics (MD) simulations often fail when trajectories sample local atomic environments that are underrepresented in the training set. Here, we employ an active-learning workflow to develop a robust neural network potential (NNP) for perfluorinated ionomer membranes (Nafion) across a wide range of hydration levels ($\lambda$). A reliable deep potential (DP) model is constructed through iterative dataset expansion within active-learning loops. Specifically, off-equilibrium configurations are generated via non-equilibrium DPMD simulations and selected using an ensemble force-deviation criterion combined with a three-dimensional structural feature space augmented by minimum interatomic distances, which significantly enhances the DP model's robustness. The trained DP model enables stable MD simulations of large Nafion systems containing approximately 10,000–20,000 atoms for an extended duration of 31 ns. Our DPMD simulations reproduce the qualitative hydration dependence of density and yield self-diffusion coefficients of hydrogen atoms and hydronium ions in quantitative agreement with experimental values. Compared with previous *ab initio* MD and MLIP-MD studies, our simulations show improved agreement with experimental transport properties and remain predictive up to $\lambda = 24$, well beyond the training range ($\lambda \leq 13$). This work provides an efficient and scalable approach for achieving stable, large-scale NNP-MD simulations of heterogeneous polymer electrolyte membranes and related disordered materials.

# 1. INTRODUCTION

Atomistic simulations using machine learning interatomic potentials (MLIPs) have garnered significant attention as a method capable of analyzing larger systems at the atomic level while maintaining accuracy comparable to electronic structure calculations.[1–3] MLIPs, trained on datasets generated from electronic structure calculations such as density functional theory (DFT) or *ab initio* wavefunction theories, offer a more accurate representation of the multidimensional potential energy surface compared to traditional classical potentials. This allows for high-accuracy, long-time molecular dynamics (MD) simulations of large systems, which are intractable with electronic structure calculations. Thus far, various MLIPs have been proposed, including kernel-based potentials,[4–7] neural network potentials (NNPs),[8–13] and more sophisticated approaches based on graph neural networks[14–22] and transformers.[23–25] Their applications span a wide range of systems, including inorganic materials,[26–30] molecular systems,[9,10,31–34] solutions,[35–38] polymeric materials,[39–42] and interfacial systems,[43,44] demonstrating the effectiveness of MLIPs across diverse systems.

While the application of MLIPs to material simulations is rapidly progressing, there are limited studies focusing on the stability of large-scale, long-time MLIP-MD simulations.[45–47] Unlike conventional classical potentials, typical atom-centered MLIPs lack physically motivated bonded/non-bonded interaction terms. Consequently, if local atomic environments underrepresented in the training dataset emerge during an MLIP-MD simulation, the prediction accuracy of atomic forces can significantly deteriorate, leading to unphysical trajectories and ultimately simulation collapse. This stability issue is particularly crucial for inhomogeneous systems such as defect-containing systems and amorphous systems, where large-scale, long-time MD simulations are essential to capture the inherent heterogeneity and slow dynamical processes of the systems. Therefore, during the generation of training data for an MLIP model, it is imperative to sample diverse structures and construct a dataset that sufficiently encompasses the local atomic environments likely to emerge during production MD runs.

We have recently developed a software package[48] for constructing a robust NNP applicable to large-scale, long-time MD simulations by generating a dataset encompassing diverse structures through active learning (AL).[49] Our NNP generator streamlines the complex steps involved in NNP construction, such as structure sampling, screening, labeling with electronic structure calculations, model training, and accuracy evaluation. Furthermore, by combining structure sampling via non-equilibrium NNP-MD simulations with subsequent screening in a structural feature space augmented by minimum interatomic distances, our NNP generator allows for efficient sampling of off-equilibrium structures, enhancing the NNP's robustness for long-time simulations.[48]

In this study, we apply this workflow to develop a robust NNP for perfluorinated ionomer membranes (Nafion) and conduct long-time, large-scale NNP-MD simulations of systems containing approximately 10,000–20,000 atoms. Because Nafion membranes exhibit highly heterogeneous morphologies that strongly depend on the hydration level ($\lambda$),[50] capturing proton-transport dynamics requires simulations that are both large in scale and long in duration. We therefore construct a lightweight Deep Potential (DP) model[51] using an iterative AL workflow and perform DPMD simulations of large Nafion systems. Subsequently, we evaluate the simulation stability based on the uncertainty in the predicted atomic forces and augment the dataset to further refine the DP model. Specifically, in the later AL loops, off-equilibrium configurations are generated via non-equilibrium DPMD simulations and selected based on an ensemble force-deviation criterion combined with a three-dimensional structural feature space augmented by minimum interatomic distances, which significantly enhances the DP model's robustness. Using the final DP model, we achieve stable MD simulations of large Nafion systems for 31 ns, reproduce the qualitative hydration dependence of density, and obtain self-diffusion coefficients of hydrogen atoms and hydronium ions in quantitative agreement with experimental values. Compared with previous *ab initio* MD (AIMD)[52,53] and MLIP-MD studies,[40] our results show improved agreement with experimental transport properties and remain predictive up to $\lambda = 24$, well beyond the training range ($\lambda \leq 13$). Overall, this work provides a practical and scalable strategy for constructing robust NNPs for heterogeneous polymer materials, where stability is often the key bottlenecks for long-time simulations.

## 2. METHODS

This section describes the methods for constructing NNP models for Nafion systems (Section 2.1) and the conditions for long-time NNP-MD simulations of large Nafion systems (Section 2.2).

**2.1. Construction of NNP models.** The process of constructing an NNP model for simulating Nafion systems is illustrated in Figure 1. Each phase of this process is described in detail below.

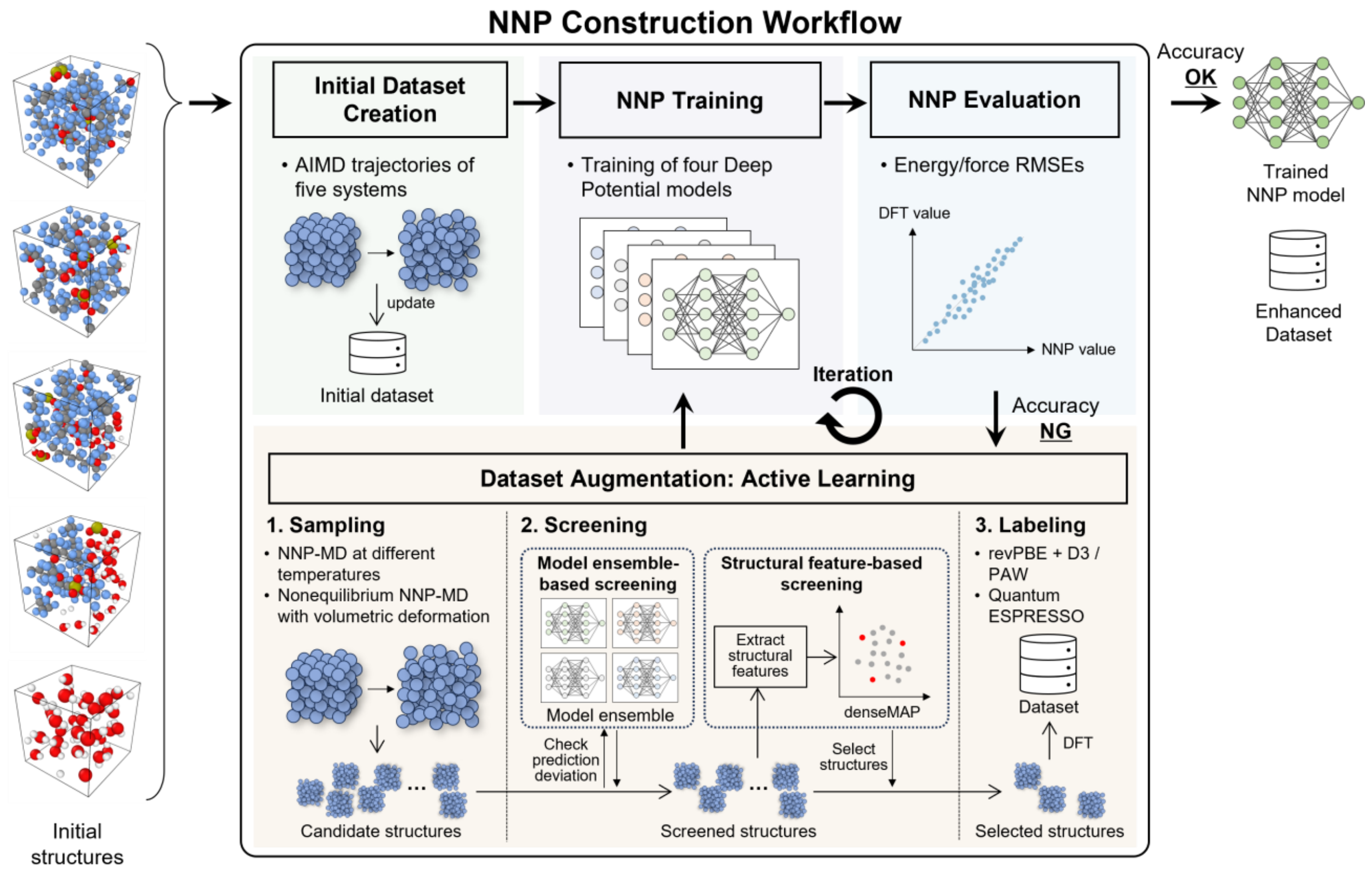

**Figure 1.** A workflow of the NNP construction process for Nafion systems.

**2.1.1. Preparation of Initial Structures.** Multiple systems comprising Nafion monomers and water molecules are used as the initial structures for NNP construction. As shown in Figure 2(a), the Nafion monomer consists of a hydrophobic Teflon backbone composed of carbon and fluorine atoms and a perfluoro side chain with a sulfonic acid group, totaling 71 atoms. The initial structures are constructed with hydration levels $\lambda = 0$, 1, 7, and 13, where $\lambda$ is defined as the number of $H_2O$ molecules per $SO_3^-$ group. These systems comprise four Nafion monomers ($\lambda = 0$), three monomers and 3 water molecules ($\lambda = 1$), three monomers and 21 water molecules ($\lambda = 7$), and two monomers and 26 water molecules ($\lambda = 13$), respectively (Figure 2(b)). Additionally, a system consisting of 32 water molecules is also included in the initial structures. RadonPy[54] is employed to generate the initial structures of the Nafion systems. Section S1 of the Supporting Information details the procedure for generating these initial structures.

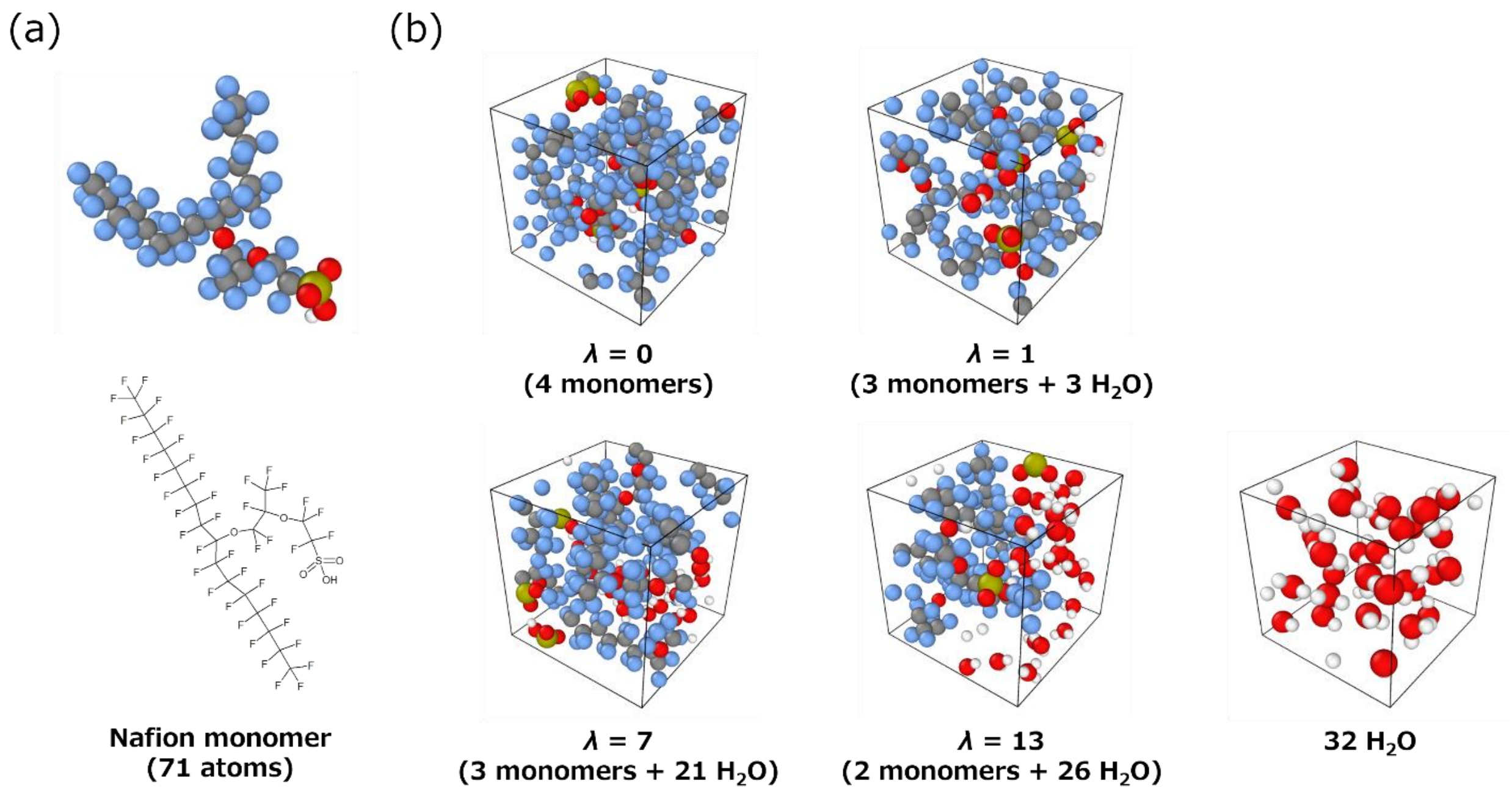

**Figure 2.** Nafion systems used for NNP training. (a) Monomer structure with 71 atoms. (b) Nafion systems and a water system. The hydration levels ($\lambda$) defined as the number of $H_2O$ molecules per $SO_3^-$ group are 0 (4 monomers), 1 (3 monomers and 3 $H_2O$ molecules), 7 (3 monomers and 21 $H_2O$ molecules), and 13 (2 monomers and 26 $H_2O$ molecules). The white, gray, red, blue, and yellow spheres represent hydrogen, carbon, oxygen, fluorine, and sulfur atoms, respectively. All the systems are visualized using OVITO.[55]

**2.1.2. Active Learning of NNPs.** As described in our previous work,[48] a single active-learning (AL) loop comprises the following steps: NNP-MD-based structure sampling, model ensemble- and structural feature-based two-stage screening, DFT-based labeling, and retraining of the NNPs with the augmented dataset. Increasing the number of AL-loop iterations enhances the structural diversity within the dataset. In this study, a total of 17 AL loops is performed, as outlined in Table 1. Furthermore, to construct the initial NNPs for the first AL loop (iteration 1), an initial dataset is generated from *ab initio* MD (AIMD) simulations and used for training the initial NNPs (iteration 0). The details of each phase are described below.

**Table 1.** The Sampling Method Used in Each AL Loop and the Numbers of Sampled, Screened, and DFT-Labeled Structures

| Iter. | Sampling | | | | 1st Screening | 2nd Screening | Labeling |
|---|---|---|---|---|---|---|---|
| | Method | Conditions[a] | Length (ps) | # structures | # structures | # structures | # structures |
| 0 | AIMD | NVT ($T$ = 400 K for Nafion, 353 K for water) | 1 | 1,000 | — | — | 1,000 |
| 1 | DPMD | NVT, NPT ($T$ = 300, 600, 900 K) | 50 | 13,707 | 2,043 | 1,000 | 784 |
| 2 | DPMD | NVT, NPT ($T$ = 300, 600, 900 K) | 50 | 22,018 | 10,293 | 1,000 | 944 |
| 3 | DPMD | NVT, NPT ($T$ = 300, 600, 900 K) | 50 | 23,727 | 9,808 | 1,000 | 995 |
| 4 | DPMD | NVT, NPT ($T$ = 300–700 K[b]) | 100 | 37,593 | 16,128 | 1,000 | 982 |
| 5 | DPMD | NVT, NPT ($T$ = 300–700 K[b]) | 100 | 39,870 | 18,659 | 1,000 | 979 |
| 6 | DPMD | NVT, NPT ($T$ = 300–700 K[b]) | 100 | 35,701 | 15,776 | 1,000 | 998 |
| 7 | DPMD | NVT, NPT ($T$ = 300–700 K[b]) | 100 | 42,076 | 25,873 | 1,000 | 981 |
| 8 | DPMD | NVT, NPT ($T$ = 300–700 K[b]) | 100 | 42,818 | 23,929 | 1,000 | 944 |
| 9 | DPMD | NVT, NPT ($T$ = 300–700 K[b]) | 100 | 43,812 | 29,374 | 1,000 | 929 |
| 10 | DPMD | NVT, NPT ($T$ = 300–700 K[b]) | 100 | 42,204 | 23,289 | 1,000 | 886 |
| 11 | NE-DPMD | Volume contraction (× 0.7, $T$ = 300–700 K[b]) | 10 | 50,000 | 31,844 | 1,000 | 993 |
| 12 | NE-DPMD | Volume contraction (× 0.7, $T$ = 300–700 K[b]) | 10 | 50,000 | 31,379 | 1,000 | 876 |
| 13 | NE-DPMD | Volume contraction (× 0.7, $T$ = 300–700 K[b]) | 10 | 50,000 | 36,827 | 1,000 | 893 |
| 14 | NE-DPMD | Volume contraction (× 0.7, $T$ = 300–700 K[b]) | 10 | 48,713 | 30,584 | 1,000 | 980 |
| 15 | NE-DPMD | Volume contraction (× 0.7, $T$ = 300–700 K[b]) | 10 | 50,000 | 36,918 | 1,000 | 998 |
| 16 | NE-DPMD | Volume contraction (× 0.7, $T$ = 300–700 K[b]) | 10 | 50,000 | 37,919 | 1,000 | 1,000 |
| 17 | NE-DPMD | Volume expansion (× 2, $T$ = 300–700 K[b]) | 10 | 50,000 | 32,510 | 1,000 | 603 |

[a]The pressures during DPMD sampling in the NPT ensemble are set to be 1 bar. [b]The temperature interval is 100 K.

***Construction of the Initial Dataset.*** To generate the initial dataset for training the initial NNPs, AIMD simulations are performed for each system shown in Figure 2(b) using Quantum ESPRESSO.[56,57] The revPBE functional[58,59] is employed with Grimme's D3 dispersion correction.[60] The pseudopotentials of the projector augmented wave (PAW) form[61] are used. The kinetic energy cutoffs for wavefunctions and charge density are set to 70 Ry and 280 Ry, respectively. The Γ point is used for the Nafion systems, while a four k-point mesh is used for the water system. The timestep size is set to 0.5 fs. All AIMD simulations are performed in the NVT ensemble for 1 ps under the periodic boundary conditions. The temperatures of the Nafion systems and the water system are maintained at 400 K and 353 K, respectively, using the Berendsen thermostat[62] with a time constant of 50 fs. Frames are stored every 50 fs, resulting in 200 frames per system, totaling 1,000 frames. The frames for each system are randomly divided into training and validation datasets with an 8:2 ratio. Each frame contains labels for total energy and atomic forces.

***NNP Training.*** DeePMD-kit 2.2.4[51] is used for NNP training. The se_e2_a descriptor of DeepPot-SE[11] is employed with a cutoff radius of 6 Å. The embedding and fitting networks consist of (25, 50, 100) and (120, 120, 120) nodes with ResNet-like architectures,[63] respectively. The number of training steps is set to 500,000 in iterations 0–16 and 2,000,000 in iteration 17. The batch size is 1. The learning rate decays exponentially from $1.0 \times 10^{-3}$ to $1.0 \times 10^{-8}$ with 5,000 decay steps. The training labels are total energy and atomic forces. The loss function is a weighted sum of energy and force loss terms, where the prefactor for each term varies linearly with respect to the learning rate. The prefactors for the energy and force loss terms at the initial training step are set to 0.02 and 1,000, respectively, and both converge to 1 as the number of training steps approaches the prescribed value. The Adam stochastic gradient descent method[64] is used to train four Deep

Potential (DP) models with independently initialized weight parameters. These four DP models are used to calculate the model deviations of atomic forces[65] in the screening phase.

***Sampling.*** To sample candidate structures to be added to the dataset, multiple DPMD simulations are performed for each system shown in Figure 2(b) using one of the four DP models generated in the previous AL loop. The conditions for the DPMD simulations in each AL loop are summarized in Table 1. In AL-loop iterations 1–10, DPMD sampling is performed in the NVT and NPT ensembles with varied temperatures, where the target pressure in the NPT ensemble is set to 1 bar. In iterations 11–16, non-equilibrium DPMD (NE-DPMD) sampling is performed by compressing the simulation cell volumes to 70% of their original size (× 0.7). Subsequently, in iteration 17, NE-DPMD sampling is conducted by expanding the simulation cell volumes to twice their original size (× 2). As discussed in Section 3.1, the NE-DPMD sampling from iteration 11 onwards aims to improve the stability of DPMD simulations by introducing off-equilibrium structures into the dataset. Specifically, the compression simulations in iterations 11–16 aim to sample structures with short interatomic distances, while the expansion simulation in iteration 17 focuses on sampling intermolecular interactions, which significantly affect the thermodynamic properties of a reference system.[47] The timestep size is set to 0.5 fs and the temperature and pressure are controlled using the Nosé–Hoover style thermostat[66,67] and barostat,[68,69] respectively. Frames are extracted from each DPMD simulation at intervals of 50 fs in AL-loop iterations 1–3, 100 fs in iterations 4–10, and 5 fs in iterations 11 and later. However, if a DPMD simulation collapses prematurely, frames are extracted only up to the point of simulation collapse. LAMMPS[70] is used to perform the DPMD simulations. Table 1 lists the number of structures sampled in each AL loop. In the initial AL-loop iterations, the accuracy of the DP model is low, leading to premature collapse in DPMD simulations and consequently fewer sampled structures. As the AL loop progresses and the DP model accuracy improves, more DPMD simulations complete successfully, resulting in an increased number of sampled structures.

***Screening.*** The structures obtained in the sampling phase undergo a two-stage screening process. The first stage involves screening based on the model deviation of predicted atomic forces defined as[51,65]

$$\epsilon_{\mathbf{F},i}(\mathbf{x}) = \sqrt{\langle \| \mathbf{F}_i(\mathbf{x};\boldsymbol{\theta}_k) - \langle \mathbf{F}_i(\mathbf{x};\boldsymbol{\theta}_k) \rangle \|^2 \rangle} \quad (1)$$

where $\mathbf{F}_i$ is a force acting on atom $i$, $\langle\cdot\rangle$ is the ensemble average over the four DP models, $\boldsymbol{\theta}_k$ is the parameters of the model $k$, and $\mathbf{x}$ denotes the coordinates and chemical species of all atoms in a certain structure. The maximum model deviation is calculated for each structure, and only those with the deviations within a specific range are selected. In this study, the lower and upper thresholds are set to 0.05 eV/Å and 0.2 eV/Å, respectively. This allows for the exclusion of structures already well-represented in the current dataset and those with very high uncertainty, which are likely to be unphysical, enabling efficient and incremental improvement of the NNP accuracy. Following the first screening, candidate structures undergo a second screening based on structural features.[71] As shown in Table 1, the number of structures selected in the first stage ranges from at least 2,000 to over 30,000 in some cases. Performing DFT calculations for all these structures is computationally prohibitive. Moreover, the selected structures often include many similar configurations in terms of structural features, resulting in redundant structural data. Consequently, the objective of the second screening is to achieve a more uniform selection of structurally diverse candidates by eliminating those exhibiting high proximity within the structural feature space. Here, the atom-wise outputs of the DP model's embedding network are concatenated and dimensionally reduced to a 2D feature space using densMAP.[72] Subsequently, screening is performed based on inter-point distances within this feature space, aiming to sample data points that are maximally distant from existing data points and ensure that the selected data points are also well-separated from each other. In AL-loop iterations 1–10 and 17, screening is performed in this 2D structural feature space (Figure S1(a)). In iterations 11–16, screening is performed in a 3D structural feature space, incorporating an additional feature related to the minimum interatomic distance (Figure S1(b)).[48] As discussed in Section 3.1, the minimum O–H distance is used as the third dimension in iterations 11–13, and the minimum F–S distance is used in iterations 14–16. For more details regarding the screening in the 3D structural feature space, please refer to our previous work.[48] Table 1 shows the number of structures selected after the model ensemble-based and structural feature-based screening. In this study, 1,000 structures are selected after the structural feature-based screening.

***Labeling.*** DFT calculations are performed for the 1,000 structures selected in the screening phase to obtain labels for total energy and atomic forces. The DFT calculation conditions are identical to those used for generating the initial dataset. For each system, the newly labeled frames are randomly divided into training and validation datasets with an 8:2 ratio. However, structures with a maximum atomic force greater than 20 eV/Å are excluded from the dataset. Table 1 shows the final number of structures labeled by DFT calculations. The discrepancy between the number of structures after screening (1,000) and that of labeled structures arises from the exclusion of structures for which DFT calculations did not converge or those exhibiting maximum atomic forces greater than 20 eV/Å.

The energy and force root-mean-square errors (RMSEs) at the end of AL loop iterations 10, 13, 16, and 17 are presented in Table S1 of the Supporting Information. In addition, parity plots comparing the DFT reference values with the DP predictions in iteration 17 for the per-atom energies and atomic forces of the Nafion and pure water systems are provided in Figure S2 of the Supporting Information. In all the iterations, the energy RMSEs on the validation set are at most around 2 meV/atom, indicating high accuracy in energy prediction. The force RMSEs are consistently larger for the Nafion systems compared to the pure water system in all iterations. This is attributed to the greater diversity of atomic environments in the five-element hydrated Nafion systems compared to the two-element water system, making the prediction of atomic forces more challenging. Nevertheless, at the end of iteration 17, the force RMSEs for the Nafion systems are at most approximately 65 meV/Å, and the force RMSE for the pure water system is below 50 meV/Å (Table S1). These results demonstrate sufficiently high accuracy in predicting atomic forces for multiple systems using a single DP model. The corresponding energy RMSEs are approximately 1 meV/atom for both the Nafion and pure water systems. These low errors indicate that the DP model accurately reproduces the underlying revPBE-D3 potential energy surface, despite employing a local environment description with a cutoff radius of 6 Å.

To assess the quality of the DFT reference forces, we analyze the net force per atom for all 16,765 structures in the dataset. Figure S3 in Supporting Information shows the distributions of the norm and the *x*-, *y*-, and *z*-components of the net force per atom. The norm of the net force per atom is well below 1 meV/Å for all structures, and the individual Cartesian components are likewise negligible. These results indicate that residual forces arising from incomplete self-consistent field (SCF) convergence[73] are insignificant in the present dataset and that the DFT forces provide reliable training targets for the DP model.

**2.2. Large-Scale NNP-MD Simulations.** Long-time MD simulations of large Nafion systems are performed using the trained DP models. Ten Nafion oligomers with a degree of polymerization of 14 (Figure 3(a)) are placed in the simulation box, along with water molecules to create the Nafion systems with $\lambda = 0, 3, 6, 12, 18$, and 24 (Figure 3(b)). The systems with $\lambda = 0, 3, 6, 12, 18$, and 24 contain 9680, 10940, 12200, 14720, 17240, and 19760 atoms, respectively. To assess the structure dependence of physicochemical properties, three independent systems are generated for each hydration level. RadonPy[54] is employed to create these initial structures. Section S1 of the Supporting Information details the procedure for generating these initial structures.

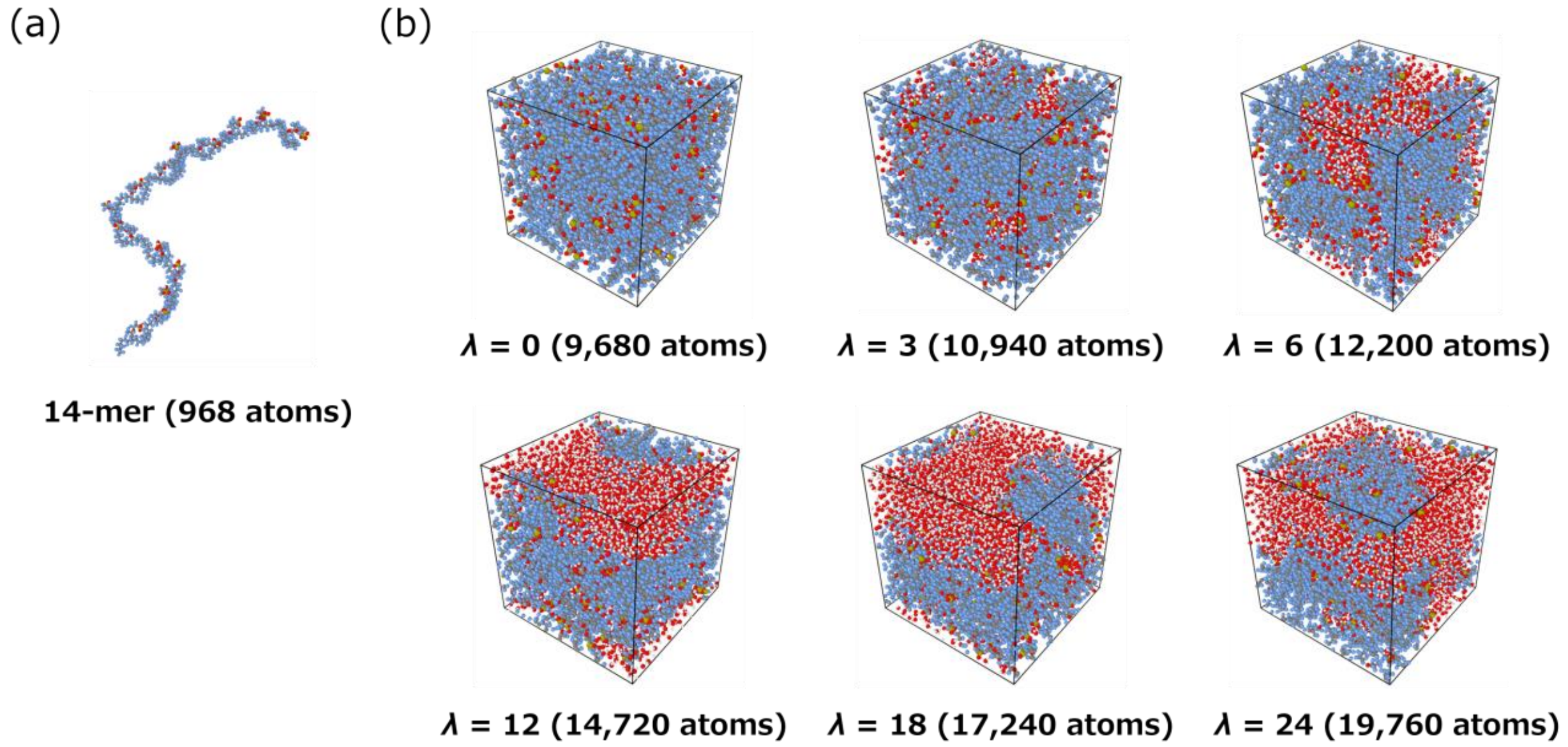

**Figure 3.** Nafion systems used for large-scale NNP-MD simulations. (a) Nafion 14-mer structure with 968 atoms. (b) Nafion systems with $\lambda = 0$ (9,680 atoms), 3 (10,940 atoms), 6 (12,200 atoms), 12 (14,720 atoms), 18 (17,240 atoms), and 24 (19,760 atoms). The number of 14-mers in each system is 10, while the number of water molecules is 0, 420, 840, 1,680, 2,520, and 3,360 for $\lambda =$ 0, 3, 6, 12, 18, and 24, respectively.

Using the generated initial structures, DPMD simulations are performed in the NPT ensemble at a temperature of 298 K and a pressure of 1 bar. The timestep size is set to 0.5 fs and the maximum simulation time is 31 ns. The temperature and pressure are controlled using the Nosé–Hoover style thermostat[66,67] and barostat[68,69] with the time constants of 50 fs and 500 fs, respectively. To evaluate the model deviations of the atomic forces at each timestep, forces acting on atoms in the system are predicted using the four NNPs and the standard deviation of these predictions is calculated.

## 3. RESULTS AND DISCUSSION

We discuss the stability of the NNP-MD simulations using DP models obtained in different iterations in Section 3.1. We show that the DP model constructed in iteration 17 enables large-scale (10,000–20,000 atoms), long-time (>30 ns) NNP-MD simulations for all the Nafion systems shown in Figure 3(b). Section 3.2 then presents hydration-dependent static properties of the Nafion systems (density and radial distribution functions) and self-diffusion coefficients of hydrogen atoms and hydronium ions calculated from the NNP-MD simulations using the DP model constructed in iteration 17. Hereafter, the DP model constructed in iteration *x* will be referred to as DP-iter-*x*.

**3.1. Improvement of Simulation Stability.** Figure 4 summarizes the stability of MD simulations of the Nafion system with $\lambda = 3$ using DP-iter-10, -13, -16, and -17 models. The MD simulation using the DP-iter-10 model exhibits a sudden decrease in the density and potential energy around 0.41 ns, indicating simulation collapse (Figure 4(a)). Correspondingly, the maximum model deviation of atomic forces also increases sharply around 0.41 ns. Although the apparent simulation collapse occurs around 0.41 ns, the maximum model deviation of atomic forces exceeds 1 eV/Å much earlier, around 0.134 ns, suggesting instability in the MD simulation. Therefore, to ensure the stability of long-time DPMD simulations, it is essential to suppress such abrupt increases in the maximum model deviation of atomic forces.

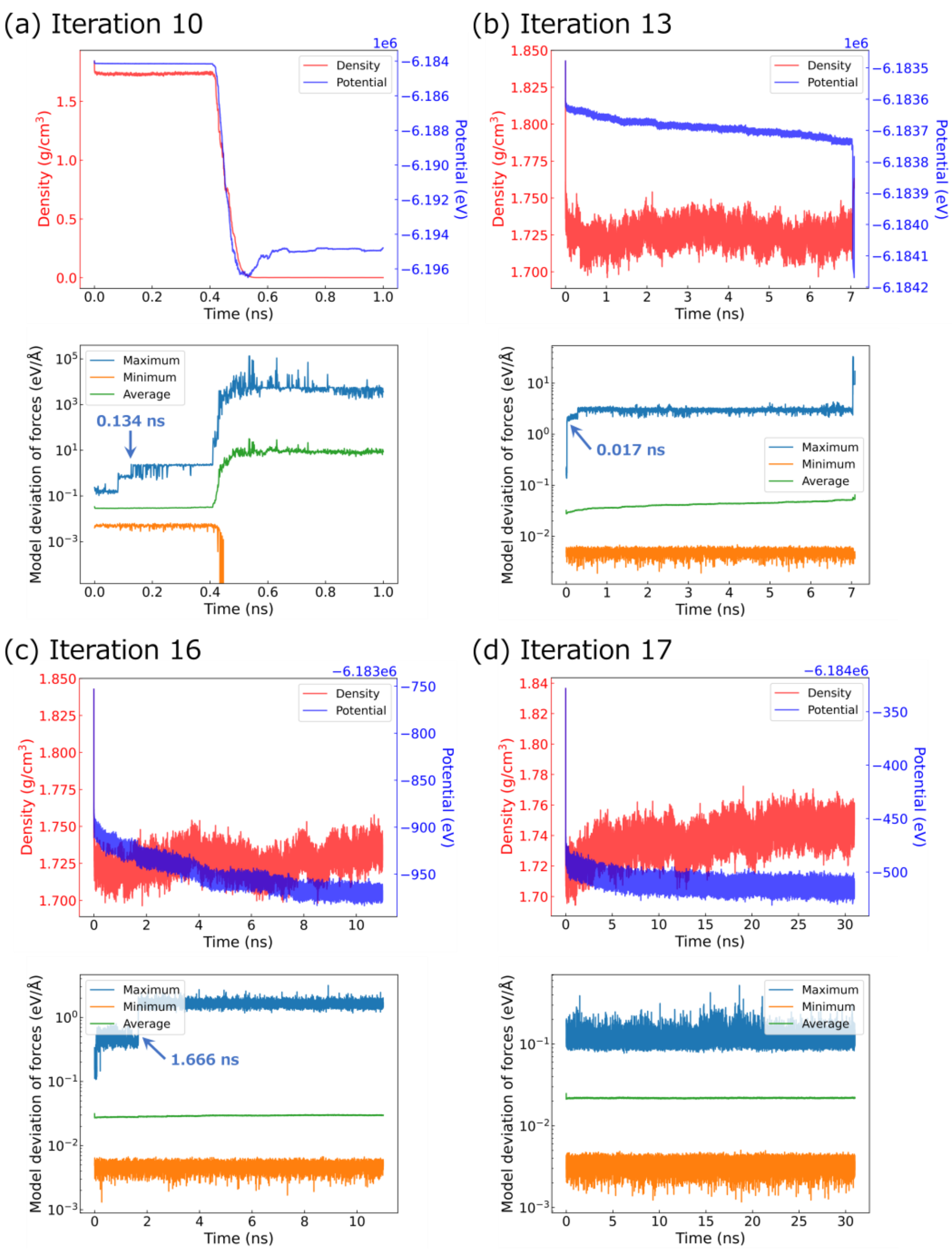

**Figure 4.** Temporal changes in the densities, potential energies, and model deviations of forces for the Nafion system with $\lambda$ = 3 simulated using the DP-iter-10 (a), -13 (b), -16 (c), and -17 (d) models. The model deviations of forces are calculated for all the atoms and their maximum, minimum, and average values are presented. The arrows in the bottom figures indicate the times when the maximum model deviations of forces exceed 1 eV/Å (see Figure 5).

Figure 5(a) compares the magnitudes and model deviations of atomic forces for the MD snapshot at $t$ = 0.134 ns using the DP-iter-10 models. As described in Section 2.2, the model deviations of the atomic forces are calculated from the predictions by the four DP-iter-10 models. The atoms exhibiting model deviations exceeding 1 eV/Å are H and O atoms. We find that the large model deviations for the H and O atoms are attributed to short O–H distances emerging in the MD simulation, of which atomic environments are underrepresented in the training dataset. To address this issue, as shown in Table 1, iterations 11–13 employ structure sampling during volume compression using NE-DPMD simulations to capture off-equilibrium structures with short interatomic distances. Furthermore, the second screening stage based on structural features incorporates the minimum O–H distance as an additional feature in a 3D structural feature space (Figure S1(b)). This combination of NE-DPMD sampling with the 3D structural feature-based screening enables more efficient selection of off-equilibrium structures with short O–H distances.[48]

To further verify this diagnosis, we compare the distributions of O–H distances sampled in iterations 0–10 and iterations 11–13 (Figure S4 of the Supporting Information). Although the total number of O–H pairs within 2.5 Å is smaller in iterations 11–13, the number of O–H pairs in the very short-distance range of 0.83–0.90 Å increases from 331 to 469. This confirms that the NE-DPMD sampling with volume compression and the 3D structural feature-based screening using the minimum O–H distance effectively enrich the short-range O–H configurations responsible for the instability of the DP-iter-10 model.

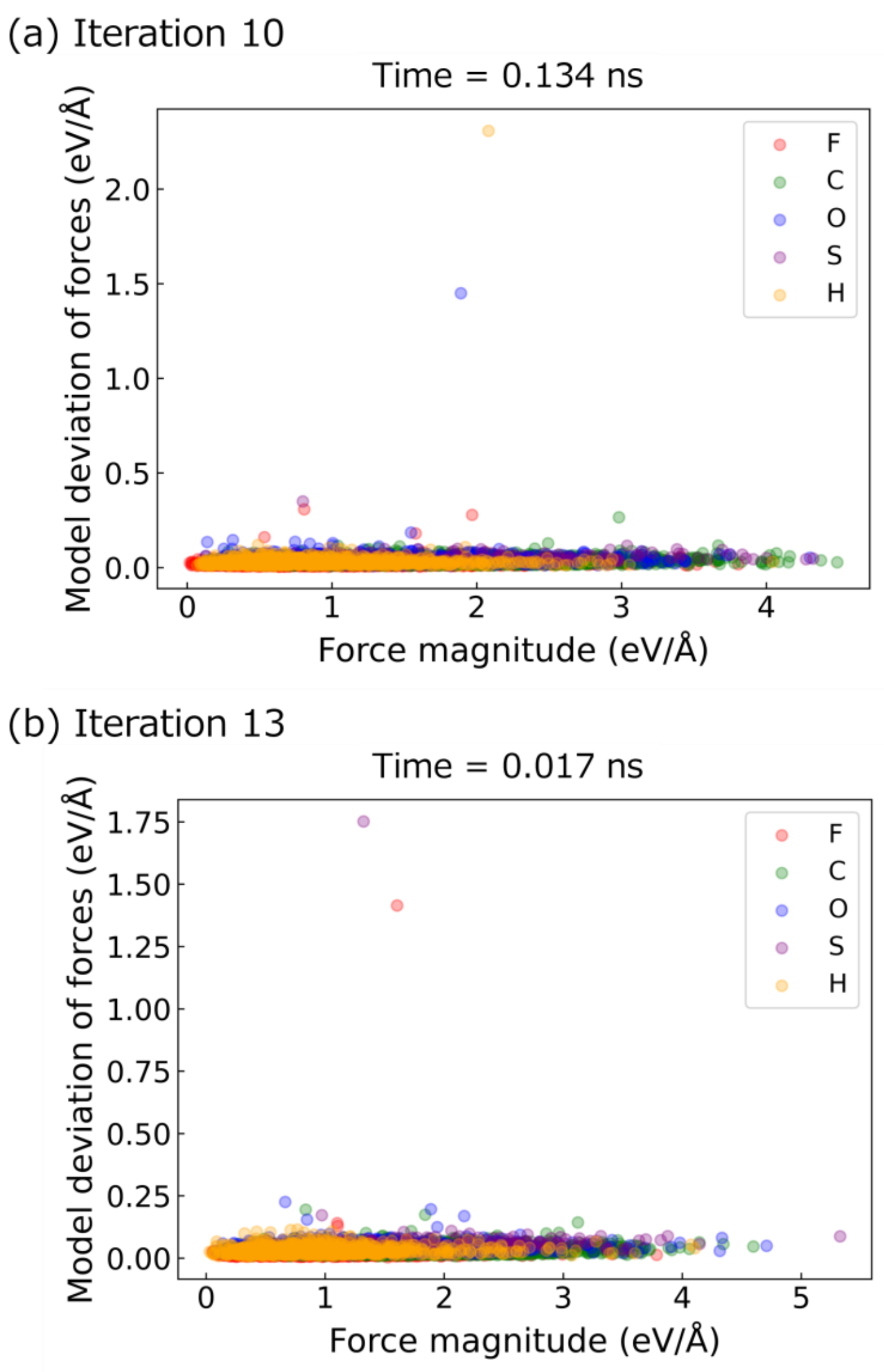

**Figure 5.** Comparison of the magnitudes and model deviations of atomic forces for the MD snapshot at $t = 0.134$ ns using the DP-iter-10 models (a) and the MD snapshot at $t = 0.017$ ns using the DP-iter-13 models (b).

As shown in Figure 4(b), the MD simulation using the DP-iter-13 model seems to be stable in terms of the density and potential energy up to around 7 ns. Subsequently, the potential energy decreases abruptly, coinciding with a sharp increase in the maximum model deviation of atomic forces, leading to simulation collapse. However, the maximum model deviation of atomic forces surpasses 1 eV/Å much earlier, around 0.017 ns, indicating underlying instability in the MD simulation. As shown in Figure 5(b), model deviations exceeding 1 eV/Å are observed for F and S atoms due to underrepresented short F–S distances in the training dataset. To address this, iterations 14–16 employ NE-DPMD simulations with volume compression for structure sampling. The minimum F–S distance is then incorporated as an additional feature in the second screening stage, enabling targeted selection of off-equilibrium structures with short F–S distances.

As shown in Figure 4(c), the MD simulation using the DP-iter-16 model seems stable in terms of the density and potential energy for over 10 ns. Although the maximum model deviation of atomic forces exceeds 1 eV/Å earlier, around 1.666 ns, the stability of the MD simulation is significantly improved compared to the simulations using the DP-iter-10 and -13 models. We find that the increase in the maximum model deviation of atomic forces is due to protons erroneously approaching the ether oxygen atoms of the Nafion oligomers in the MD simulation using the DP-iter-16 model. The structures including such atomic environments are not present in the dataset because it is almost impossible to converge DFT calculations of such unphysical structures. To address this issue, the Ziegler–Biersack–Littmark (ZBL) repulsive potential,[74] with the inner and outer cutoffs of 0.1 Å and 2 Å, respectively, is introduced between hydrogen atoms and the ether oxygen atoms of the Nafion oligomer. Furthermore, in iteration 17, structure sampling is performed during volume expansion using NE-DPMD simulations to sample a wider range of intermolecular interactions (Table 1).

As shown in Figure 4(d), the MD simulation using the DP-iter-17 model remains stable for an extended duration of 31 ns. Furthermore, the maximum model deviation of atomic forces remains on the order of $10^{-1}$ eV/Å, indicating the stability of the MD simulation. As shown in Figure S5 of the Supporting Information, the MD simulations for all other hydration levels also remain stable for durations of 31 ns. These results demonstrate the successful construction of an NNP capable of performing large-scale, long-time MD simulations. The following discussion focuses on the results obtained from MD simulations using the DP-iter-17 model.

**3.2. Evaluation of Physical Properties.** We first examine the static properties of the simulated Nafion systems. Figure 6(a) compares the densities of the Nafion systems obtained from our DPMD simulations with the experimental values.[75] Although the DPMD-calculated densities are slightly lower than the experimental values, the qualitative dependence on the hydration level is accurately captured. The minimal standard deviations observed across three independent systems underscore the statistical convergence and the robustness of our large-scale, long-time trajectories. We hypothesize that this systematic underestimation of density is partially attributable to the choice of the revPBE functional[58,59] used for the DFT-based labeling. This inference is consistent with prior research; for example, a previous study[35] reported a similar underestimation of water density (approximately 10%) when using NNP models trained on DFT data generated with the RPBE functional[76] and Grimme's D3 dispersion correction.[60]

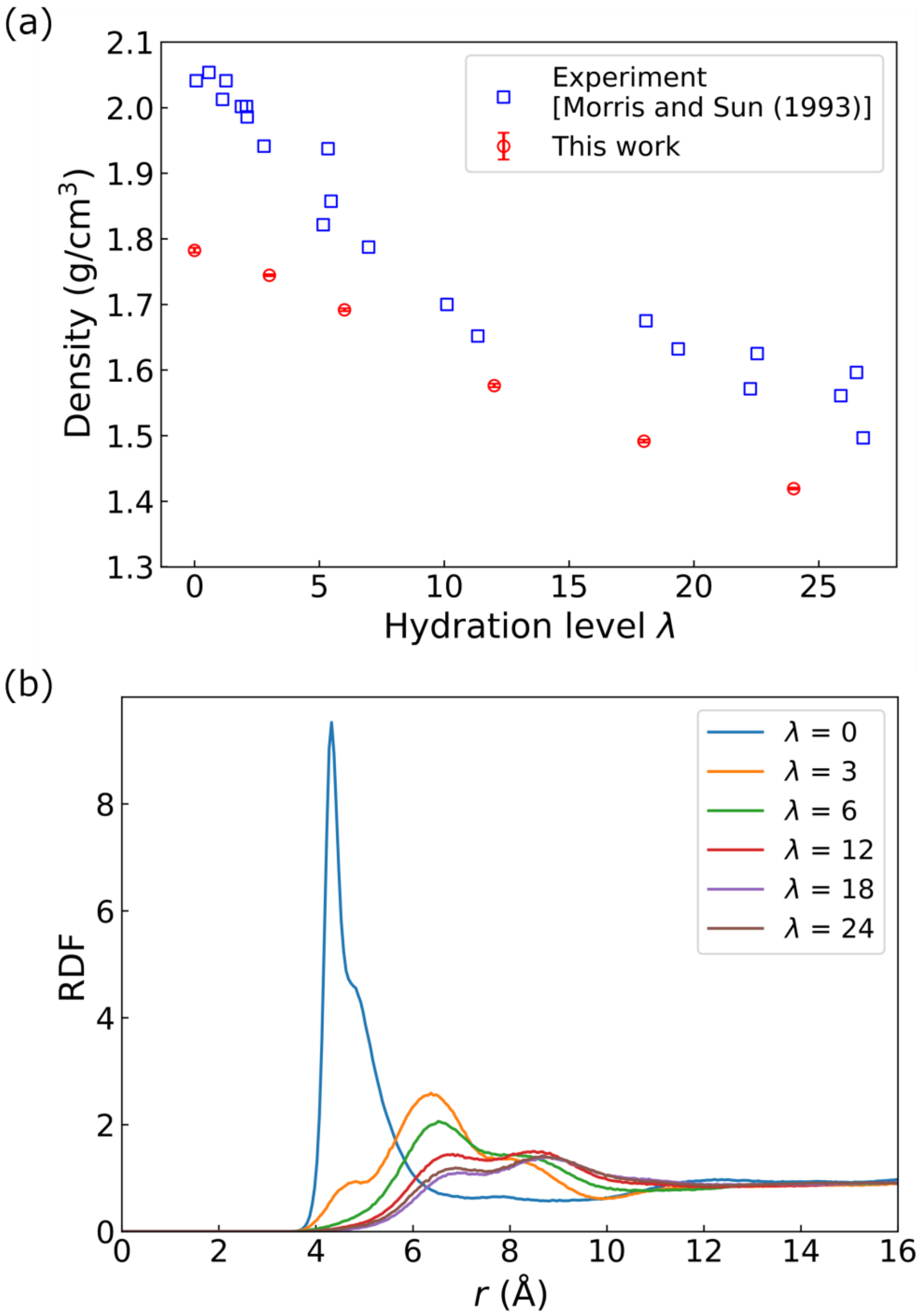

**Figure 6.** (a) Comparison of the MD-calculated densities of the hydrated Nafion systems with the experimental counterparts.[75] The error bars of the calculated densities are the standard deviations obtained from three independent simulations starting from different structures. (b) Radial distribution functions (RDFs) between sulfur atoms as a function of the hydration level.

Figure 6(b) presents the radial distribution functions (RDFs) between sulfur atoms as a function of the hydration level. In the dry membrane ($\lambda = 0$), the RDF exhibits a pronounced and sharp first peak at short distances ($r \approx$ 4–5 Å), indicating strongly correlated, tightly packed sulfonate headgroup arrangements. Upon hydration ($\lambda$ = 3–12), this first-shell feature substantially decreases in magnitude, broadens, and shifts to larger separations ($r \approx$ 6 Å), consistent with swelling of ionic domains and the intercalation of water molecules that reduce direct headgroup–headgroup contacts and promote solvent-mediated configurations. Notably, at $\lambda = 3$ we observe a weak shoulder at $r \approx$ 4–5 Å, close to the dominant first-shell distance in the dry membrane ($\lambda = 0$). This feature suggests the coexistence of a residual "contact" population of tightly associated ionic headgroups with an emerging solvent-separated population whose main peak appears at larger distances ($r \sim$ 6 Å). Concurrently, a broader second-shell feature emerges around $r \approx$ 7–9 Å, suggesting a reorganization toward a more extended nanophase-separated morphology in which hydrated ionic domains are separated by water-rich regions. At higher hydration ($\lambda \geq 18$), the RDF curves become nearly indistinguishable, implying that the average headgroup spatial organization has largely saturated. Overall, the RDF trends corroborate the monotonic density decrease in Figure 6(a) and provide a structural signature of hydration-induced swelling, heterogeneous hydration at low $\lambda$, and the eventual saturation of the ionic nanodomains at high $\lambda$.

We next quantify the extent of proton dissociation from the sulfo groups to form hydronium ions as a function of the hydration level $\lambda$. The deprotonated sulfo groups and hydronium ions are identified by assigning each hydrogen atom to its nearest oxygen atom,[77] providing a consistent and well-defined criterion for proton localization. At $\lambda = 3$, approximately 92% of the sulfo groups are deprotonated, indicating that proton dissociation is already strongly favored even under relatively low hydration conditions (Table S2 of the Supporting Information). With increasing hydration, the fraction of deprotonated sulfo groups increases further, and for $\lambda \geq 6$ nearly complete deprotonation is observed. This monotonic trend with hydration level is physically reasonable, as enhanced solvation at higher $\lambda$ facilitates proton separation and stabilization of hydronium ions.

Experimental FT-IR studies of perfluorosulfonic acid membranes have shown that protons can dissociate from sulfonic acid groups even in the presence of small amounts of water. Giffin et al.[78] reported that the vibrational band associated with a covalently bound S–OH proton disappears under low-humidity conditions, while bands associated with $SO_3^-$ groups appear, indicating proton dissociation at approximately $\lambda = 1$. In addition, theoretical studies[79,80] distinguish between hydronium–sulfonate ion-pair formation at low hydration levels and the formation of solvent-separated cations at higher hydration levels. These observations support the hydration-dependent increase in proton dissociation observed in the present simulations. Nevertheless, the absolute protonation ratios should be interpreted within the accuracy of the underlying revPBE-D3 reference calculations.

Building on the hydration-dependent proton dissociation behavior described above, we characterize the diffusive behavior of hydrogen atoms and hydronium ions by calculating their mean-squared displacements (MSDs) and evaluating the self-diffusion coefficients using the Einstein relationship,[81]

$$D = \frac{1}{6} \lim_{t \to \infty} \frac{\langle |\mathbf{r}_i(t) - \mathbf{r}_i(0)|^2 \rangle}{t} \tag{2}$$

where $\mathbf{r}_i$ is the position of particle $i$. The MSDs of hydrogen atoms in the hydrated Nafion systems are shown in Figure S6(a) of the Supporting Information. The diffusivity of hydrogen atoms clearly increases with increasing hydration level, reflecting enhanced proton mobility in more hydrated environments. Owing to the large system size and long simulation times employed in our DPMD simulations, the standard deviations of the MSDs are significantly reduced, and the diffusive regime—where the MSD increases linearly with time—is clearly captured.

Figure 7(a) presents the self-diffusion coefficients of hydrogen atoms obtained from the linear slopes of the MSDs using Equation 2. The self-diffusion coefficients predicted by our DPMD simulations are in close agreement with experimental values[82,83] over a wide range of hydration levels. In contrast, previous AIMD simulations fail to quantitatively reproduce the experimental diffusion coefficients, primarily due to limitations in system size and simulation time.[52,53] While MD simulations based on a Gaussian approximation potential (GAP)[40] successfully reproduce the diffusion coefficients at $\lambda = 3$ and 6, they underestimate the diffusion coefficient at $\lambda = 12$. This discrepancy likely arises from the relatively small system size (approximately 1,000 atoms) and short simulation duration (approximately 200 ps) employed in the GAP-MD simulations, which are insufficient to fully capture the transport behavior of hydrogen atoms in highly heterogeneous hydrated Nafion systems at higher hydration levels. In contrast, our DPMD simulations, performed for up to 31 ns on systems containing more than 10,000 atoms, accurately reproduce the hydrogen transport properties across all hydration levels investigated.

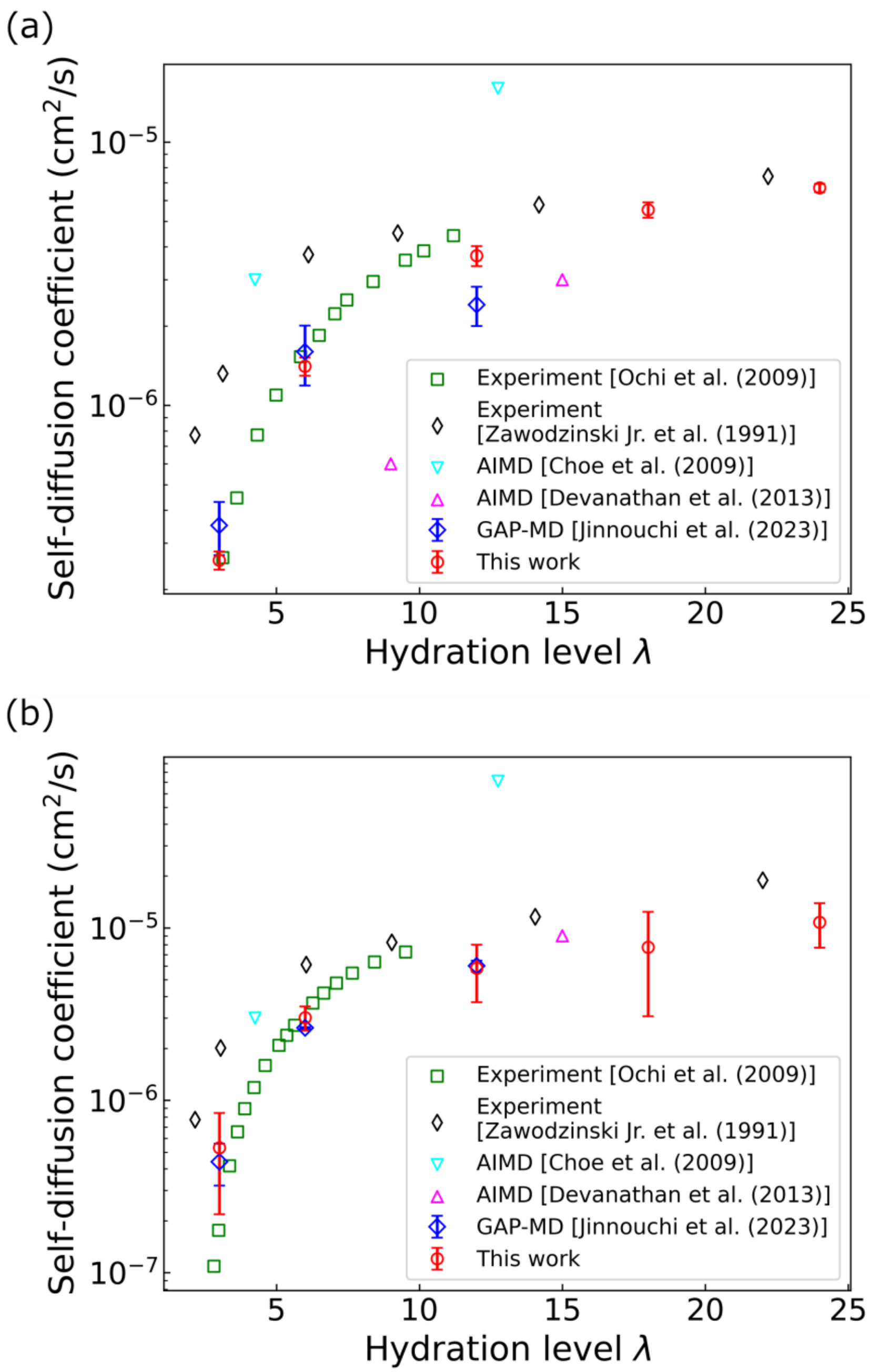

**Figure 7.** Comparison of the calculated self-diffusion coefficients of hydrogen atoms (a) and hydronium ions (b) in the hydrated Nafion systems with the experimental,[82,83] AIMD-calculated,[52,53] and GAP-MD-calculated[40] values. The error bars of the calculated self-diffusion coefficients are the standard deviations obtained from three independent simulations starting from different structures.

To examine the relaxation of slower polymer degrees of freedom, we calculate the orientational autocorrelation function[84] of local Nafion backbone segment vectors. As shown in Figure S7 of the Supporting Information, the backbone orientational autocorrelations decay slowly and do not fully decorrelate within the time window used for the MSD analysis, indicating that the global polymer conformation is not fully decorrelated on this time scale. Therefore, the present self-diffusion coefficients should be interpreted as transport properties within the equilibrated hydrated Nafion morphologies sampled in the simulations, rather than as values averaged over the complete polymer conformational ensemble. The convergence of the reported self-diffusion coefficients is assessed from the linearity of the MSDs, as shown in Figure S6 of the Supporting Information.

The MSDs of hydronium ions are shown in Figure S6(b) of the Supporting Information. Similar to hydrogen atoms, the MSDs of hydronium ions increase with increasing hydration level, indicating enhanced ionic mobility in more hydrated systems. Compared to hydrogen atoms, however, the MSDs of hydronium ions exhibit larger standard deviations. This increased fluctuation can be attributed to the smaller number of hydronium ions in the system (approximately 130–140 ions; Table S2) as well as to the discontinuous changes in their effective positions induced by proton hopping events associated with the Grotthuss mechanism.[85] Nevertheless, a linear regime in the MSDs is observed at sufficiently long times (2–3 ns), allowing reliable estimation of the self-diffusion coefficients using Equation (2). Figure 7(b) shows the resulting self-diffusion coefficients of hydronium ions. Excellent agreement with experimental values is obtained over a wide range of hydration levels. At hydration levels $\lambda$ = 3, 6, and 12, the present results are also consistent with those reported in previous GAP-MD simulations.[40]

Remarkably, our DPMD simulations accurately reproduce the self-diffusion coefficients of hydrogen atoms and hydronium ions even at hydration levels up to $\lambda = 24$ in the production runs, even though the training dataset only includes systems with hydration levels up to $\lambda = 13$ (Figure 2(b)). This successful extrapolation beyond the training domain is attributed to the diversity of local atomic environments captured in the training dataset. In particular, the active-learning procedure enables efficient sampling of a wide range of structural configurations, resulting in a dataset rich in diverse local environments. Consequently, the DP model developed in this study reliably reproduces the transport properties of hydrogen atoms and hydronium ions even outside the training range of hydration levels.

Finally, we note that no finite-size correction[86] is applied to the reported self-diffusion coefficients. This correction is appropriate for homogeneous and isotropic bulk liquids with a well-defined shear viscosity and characteristic box length. In the present hydrated Nafion membranes, however, hydrogen and hydronium diffusion occurs in a heterogeneous and anisotropic environment associated with the phase-separated hydrophilic domains and nanoscale water-channel networks. Because an effective viscosity and hydrodynamic length scale are not uniquely defined for these systems, direct application of the bulk-liquid correction would require system-specific assumptions. The reported self-diffusion coefficients are therefore left uncorrected.

## 4. CONCLUSIONS

We have developed a robust deep potential (DP) model for perfluorinated ionomer membranes (Nafion) using an iterative active-learning (AL) workflow. The key to achieving stability in long-time, large-scale simulations is systematic dataset expansion guided by uncertainty in the predicted atomic forces, together with strategic sampling of diverse configurations underrepresented in the training set. In particular, the DP model's robustness is significantly improved by generating off-equilibrium structures from non-equilibrium DP-based molecular dynamics (MD) simulations, selecting them using an ensemble force-deviation criterion, and then screening them in a three-dimensional structural feature space augmented with minimum interatomic distances.

The resulting DP model enables stable MD simulations of heterogeneous Nafion systems containing approximately 10,000–20,000 atoms for 31 ns across a broad range of hydration levels ($\lambda$). These simulations capture the qualitative hydration dependence of density and yield self-diffusion coefficients for hydrogen atoms and hydronium ions that are quantitatively consistent with experimental measurements. Notably, the DP model remains predictive up to $\lambda = 24$, substantially beyond the training range ($\lambda \leq 13$), and shows improved agreement with experimental transport properties compared with prior *ab initio* MD[52,53] and MLIP-MD[40] studies.

Overall, this work provides a scalable route to constructing robust neural network potentials for heterogeneous polymer membranes, and the proposed AL selection strategy should be transferable to other disordered polymeric and ionic materials.

## ASSOCIATED CONTENT

### Supporting Information

Additional computational details for constructing Nafion systems; representative feature spaces used for structural feature-based screening; parity plots of DP-iter-17-predicted and DFT reference per-atom energies and atomic forces; per-atom net force distributions for the DFT dataset; O–H distance distributions sampled during active-learning iterations 0–10 and 11–13; stability evaluation of long-time DP-iter-17-based MD simulations across hydration levels; mean-squared displacements of hydrogen atoms and hydronium ions; orientational autocorrelation functions of local Nafion backbone segment vectors; energy and force RMSEs for representative active-learning iterations; fractions of deprotonated sulfo groups at different hydration levels (PDF)

## AUTHOR INFORMATION

### Corresponding Author

*E-mail: yoshimoto.yuta@fujitsu.com

### Author Contributions

The manuscript was written through contributions of all authors. All authors have given approval to the final version of the manuscript.

### Notes

The authors declare no competing financial interest.

# Supporting Information

# Active Learning of a Neural Network Potential for Large-Scale Atomistic Simulations of Polymer Electrolyte Membranes

*Yuta Yoshimoto,* Naoki Matsumura, Meguru Yamazaki, Yuto Iwasaki, Hiroshi Nakao, and Yasufumi Sakai*

Fujitsu Research, Fujitsu Limited, 4-1-1 Kamikodanaka, Nakahara-ku, Kawasaki, Kanagawa 211-8588, Japan

## S1. Methods for Constructing Nafion Systems

This section describes the procedure for generating initial structures for the Nafion systems. Section S1.1 details how to construct initial structures for NNP training, while Section S1.2 describes how to construct initial structures for large-scale simulations. The Nafion systems are generated using RadonPy.[1]

### S1.1. Construction of Initial Structures for NNP Training.

***Initial Structure of the Dry Nafion System.*** Initially, 1,000 3D conformations of the Nafion monomer are generated from its SMILES string using RDKit. These conformations are then subjected to molecular mechanics optimization using the modified GAFF2 force field parameterized for fluorocarbons.[2] Subsequently, density functional theory (DFT) calculations using the PSI4 software[3] are performed to optimize the four most stable structures obtained from the molecular mechanics calculations. These DFT calculations employ the ωB97M-D3BJ functional[4,5] and the 6-31G(d,p) basis set. For the lowest energy structure, a single-point calculation using the Hartree–Fock method with the 6-31G(d) basis set is performed to obtain restrained electrostatic potential (RESP) charges.[6] Four Nafion monomers, assigned with the derived atomic charges and the modified GAFF2 parameters, are then placed in a simulation box with a target density of 0.2 g/cm$^3$. A heating simulation is performed in the NVT ensemble from 400 K to 700 K over 1 ns, using a 1-fs timestep size and the Nosé–Hoover thermostat[7,8] with a time constant of 100 fs. All bond lengths and angles involving hydrogen atoms are constrained using the RATTLE algorithm.[9] The system is then isotropically compressed to a density of 0.8 g/cm$^3$ over 1 ns at 700 K in the NVT ensemble. This is followed by a 21-step compression/decompression equilibration run[10] for 1.5 ns. Finally, a 20-ns equilibration is performed in the NPT ensemble at 400 K and 1 bar, using the Nosé–Hoover style thermostat[7,8] and barostat[11,12] with time constants of 100 fs and 1000 fs, respectively.

***Initial Structures of the Hydrated Nafion Systems.*** Modified GAFF2 parameters[2] and RESP charges are assigned to the Nafion monomers using the same procedure as that employed for the dry Nafion system. For the hydrated systems with the hydration levels $\lambda = 1$ and 7, three Nafion monomers are placed in a simulation box with a target density of 0.2 g/cm$^3$, followed by the random insertion of 3 and 21 water molecules, respectively. For the system with $\lambda = 13$, two Nafion monomers are placed in the box at a target density of 0.2 g/cm$^3$, and 26 water molecules are randomly inserted. The TIP3P model[13] is used for water. The system is then isotropically compressed to a density of 1.0 g/cm$^3$ over 5 ns at 700 K in the NVT ensemble. A 21-step compression/decompression equilibration[10] is then performed for 1.5 ns. Finally, a 20-ns equilibration is conducted in the NPT ensemble at 400 K and 1 bar.

***Initial Structure of the Water System.*** Thirty-two water molecules are randomly placed in a simulation box with a target density of 0.5 g/cm$^3$. The TIP3P model[13] is employed for water. The system is then cooled from 1000 K to 353 K over 1 ns in the NVT ensemble using a 1-fs time step and the Nosé–Hoover thermostat with a time constant of 100 fs. The O–H bond lengths and H–O–H bond angles are constrained using the RATTLE algorithm.[9] Subsequently, a 1-ns equilibration is performed in the NPT ensemble at 353 K and 1 bar, using the Nosé–Hoover style thermostat and barostat with time constants of 100 fs and 1000 fs, respectively.

### S.1.2. Construction of Initial Structures for Large-Scale Simulations.

***Initial Structure of the Dry Nafion System.*** RESP charges are assigned to the Nafion monomer following the procedure detailed in Section S.1.1. Subsequently, a Nafion 14-mer is generated using the polymerization functionality of RadonPy[1] and assigned with the modified GAFF2 parameters.[2] Ten 14-mers are then randomly placed in a simulation box, targeting a density of 0.2 g/cm$^3$. The subsequent equilibration procedure is identical to that described in Section S.1.1.

***Initial Structures of the Hydrated Nafion Systems.*** The procedure for placing ten Nafion 14-mers in the simulation box is identical to that employed for the dry Nafion system. Subsequently, 420, 840, 1,680, 2,520, and 3,360 water molecules are randomly inserted into the simulation boxes to generate hydrated systems with $\lambda = 3$, 6, 12, 18, and 24, respectively. The subsequent equilibration procedure is identical to that detailed in Section S.1.1.

## S2. Supplementary Figures and Tables to Main Text

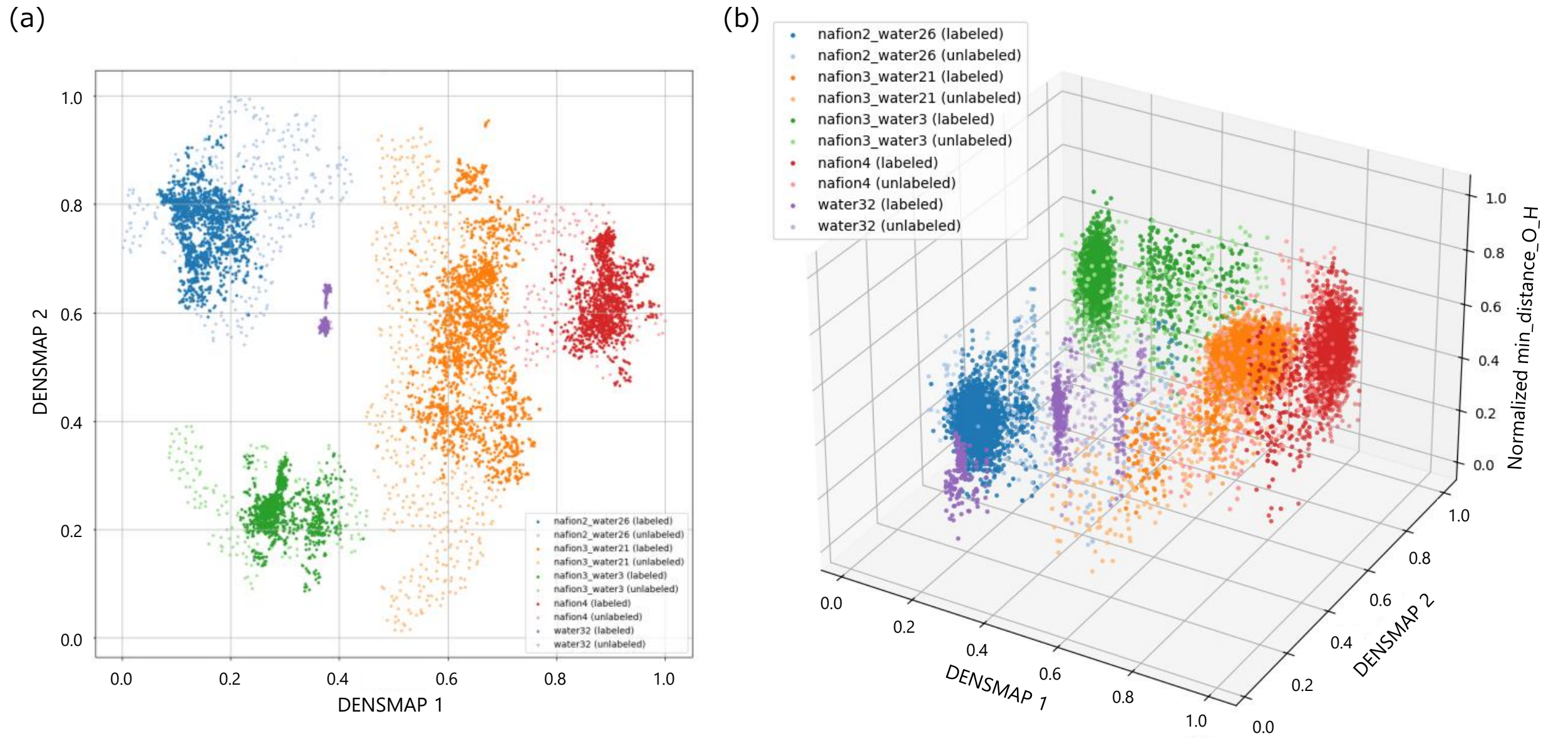

**Figure S1.** Representative two-dimensional (a) and three-dimensional (b) structural feature spaces used for structural feature-based screening. In panel (a) (iteration 10), different colors denote different systems; darker points represent the complete set of structures labeled up to (and including) iteration 9, whereas lighter points indicate structures selected for labeling in iteration 10. Panel (b) shows the feature space at iteration 13 using the same convention: darker points correspond to all structures labeled up to iteration 12, and lighter points to structures selected for labeling in iteration 13. The normalized minimum O–H distance is used as the third dimension in (b).

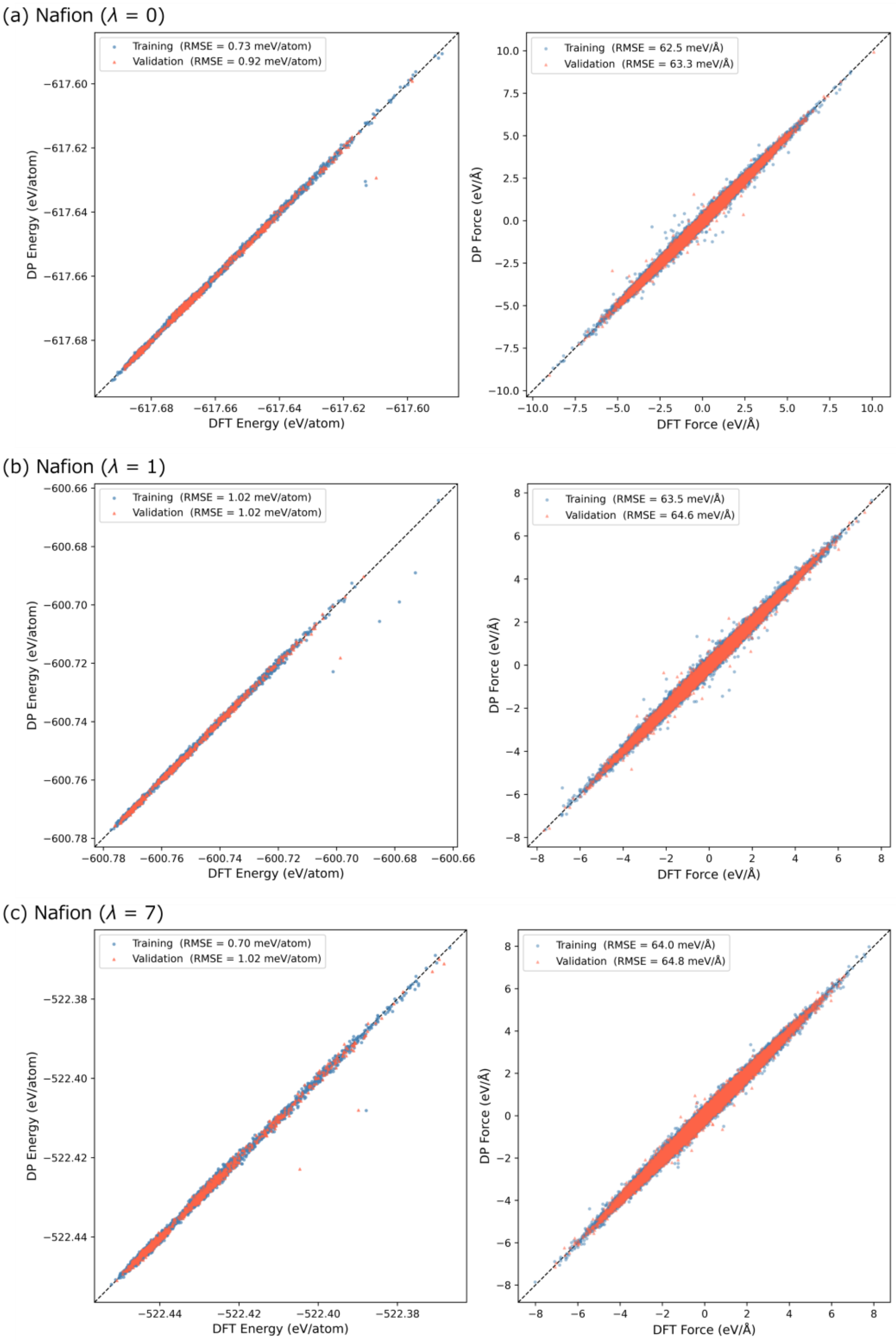

**Figure S2.** Parity plots comparing the per-atom energies (left) and atomic forces (right) predicted by the DP model in iteration 17 with the corresponding DFT reference values for the training and validation datasets. (a) Nafion ($\lambda = 0$), (b) Nafion ($\lambda = 1$), and (c) Nafion ($\lambda = 7$).

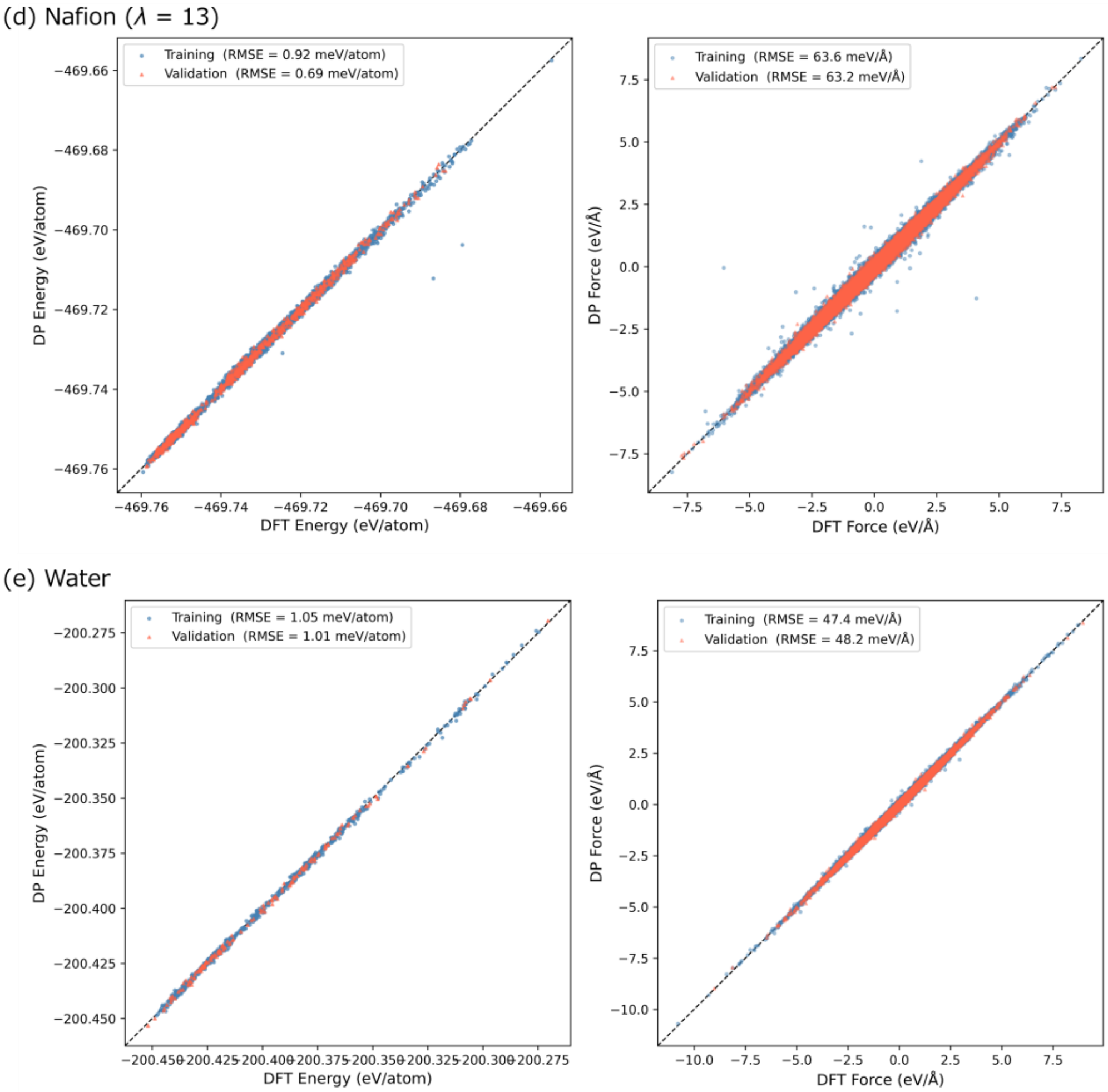

**Figure S2 (continued).** Parity plots comparing the per-atom energies (left) and atomic forces (right) predicted by the DP model in iteration 17 with the corresponding DFT reference values for the training and validation datasets. (d) Nafion ($\lambda$ = 13) and (e) water.

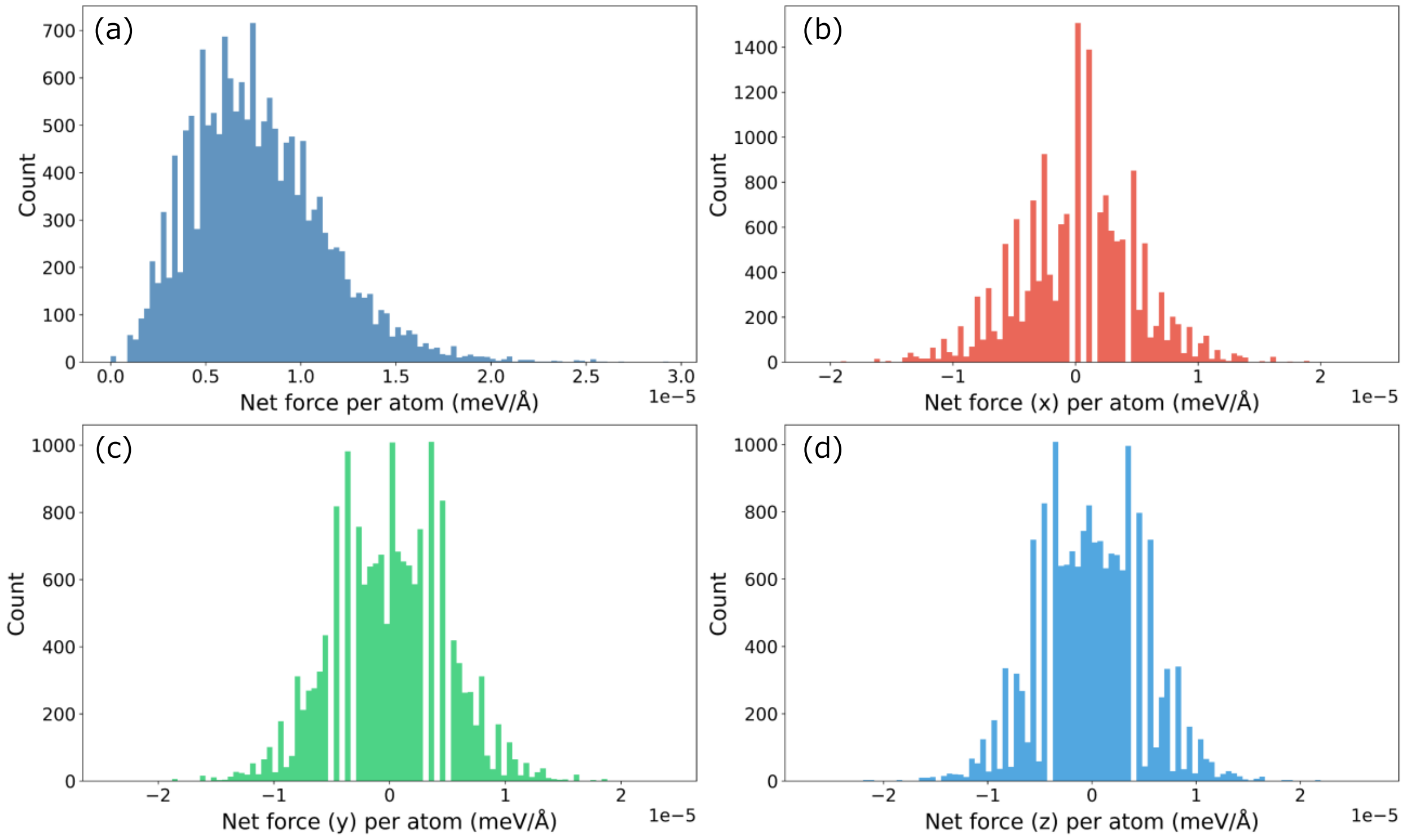

**Figure S3.** Distributions of the net force per atom for all 16,765 structures included in the DFT dataset. (a) Norm of the net force per atom. (b–d) *x*-, *y*-, and *z*-components of the net force per atom, respectively. The net force per atom is calculated by dividing the total net force of each structure by the number of atoms.

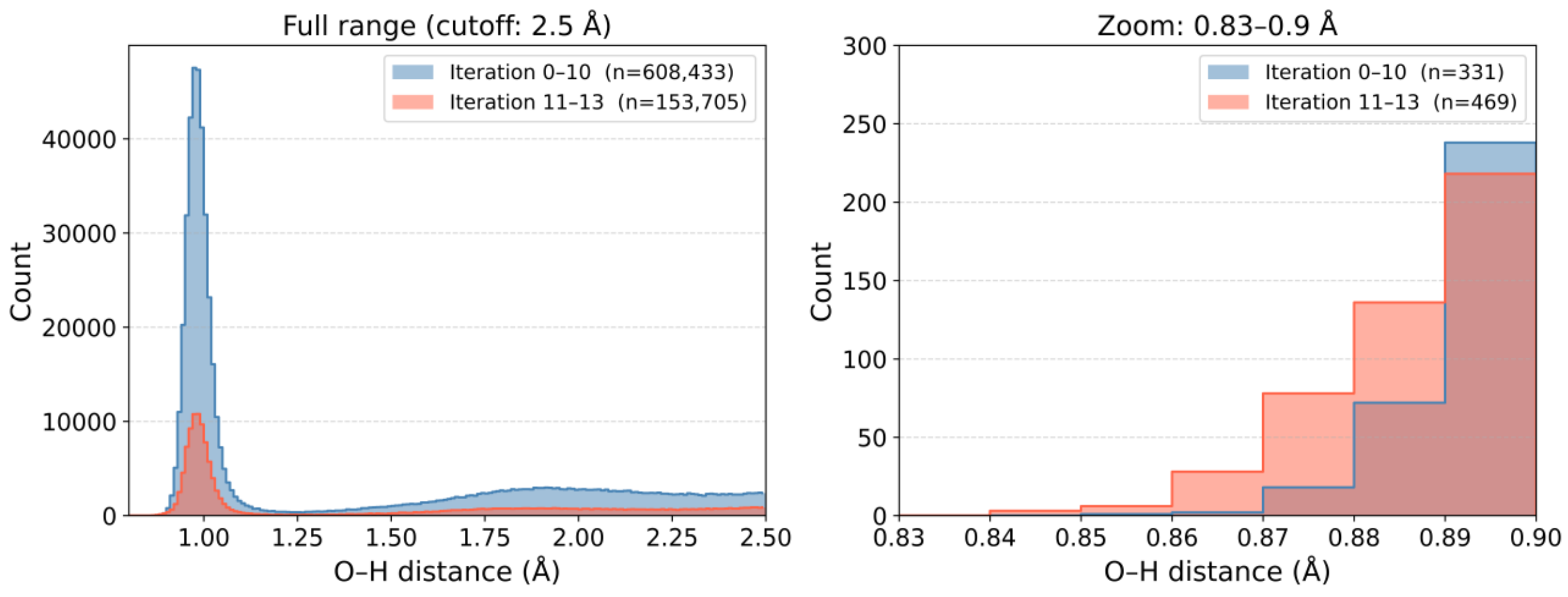

**Figure S4.** Distributions of O–H distances sampled in iterations 0–10 and iterations 11–13. The left panel shows the full distribution for O–H distances below 2.5 Å, and the right panel shows a magnified view of the short-distance region from 0.83 to 0.90 Å. The increased number of short O–H distances in iterations 11–13 indicates that the NE-DPMD sampling with volume compression and the 3D structural feature-based screening using the minimum O–H distance effectively enrich short-range repulsive configurations.

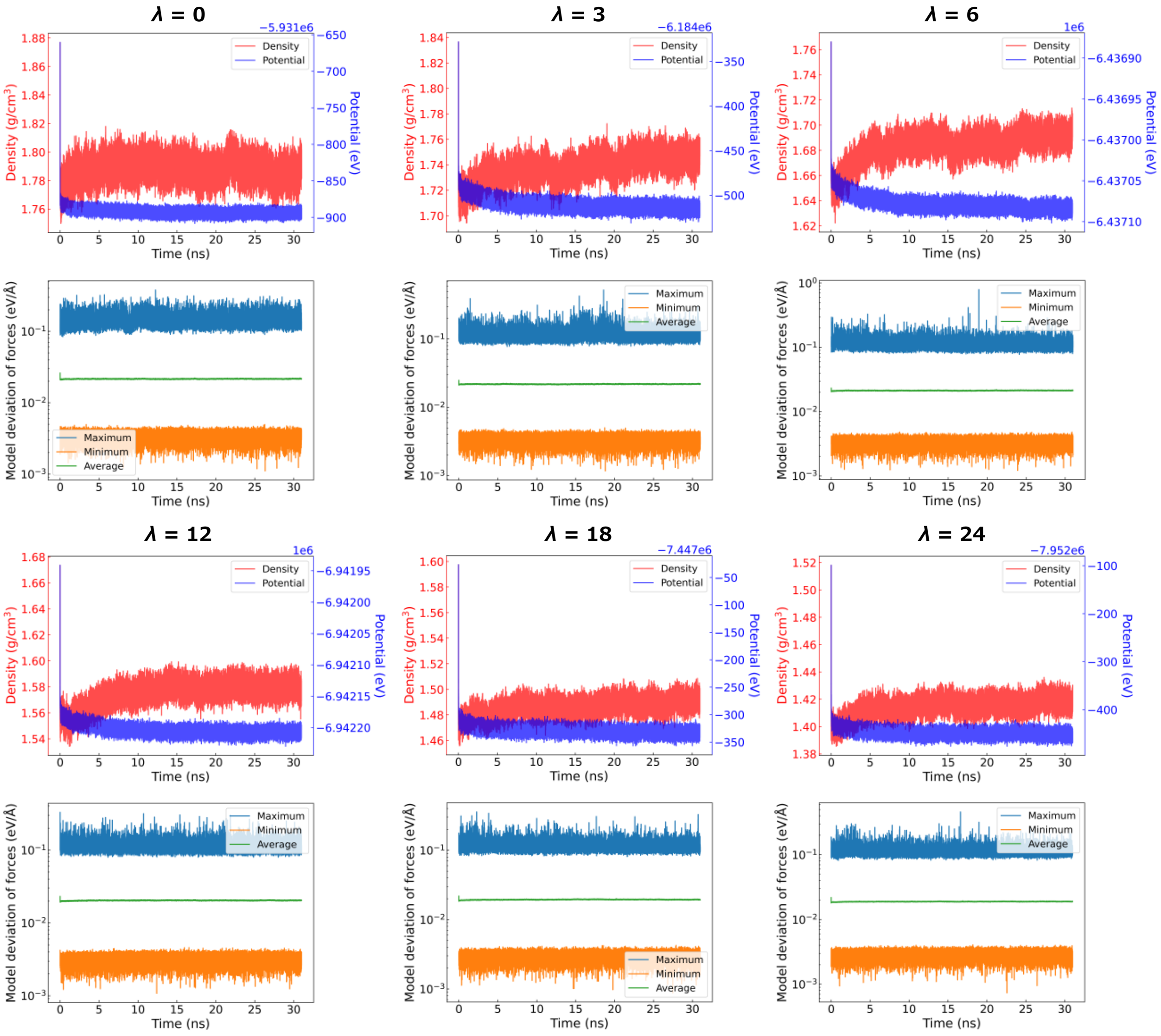

**Figure S5.** Temporal changes in the densities, potential energies, and model deviations of forces for the Nafion systems with $\lambda$ = 0, 3, 6, 12, 18, and 24 simulated using the DP-iter-17 model. The model deviations of forces are calculated for all the atoms and their maximum, minimum, and average values are presented.

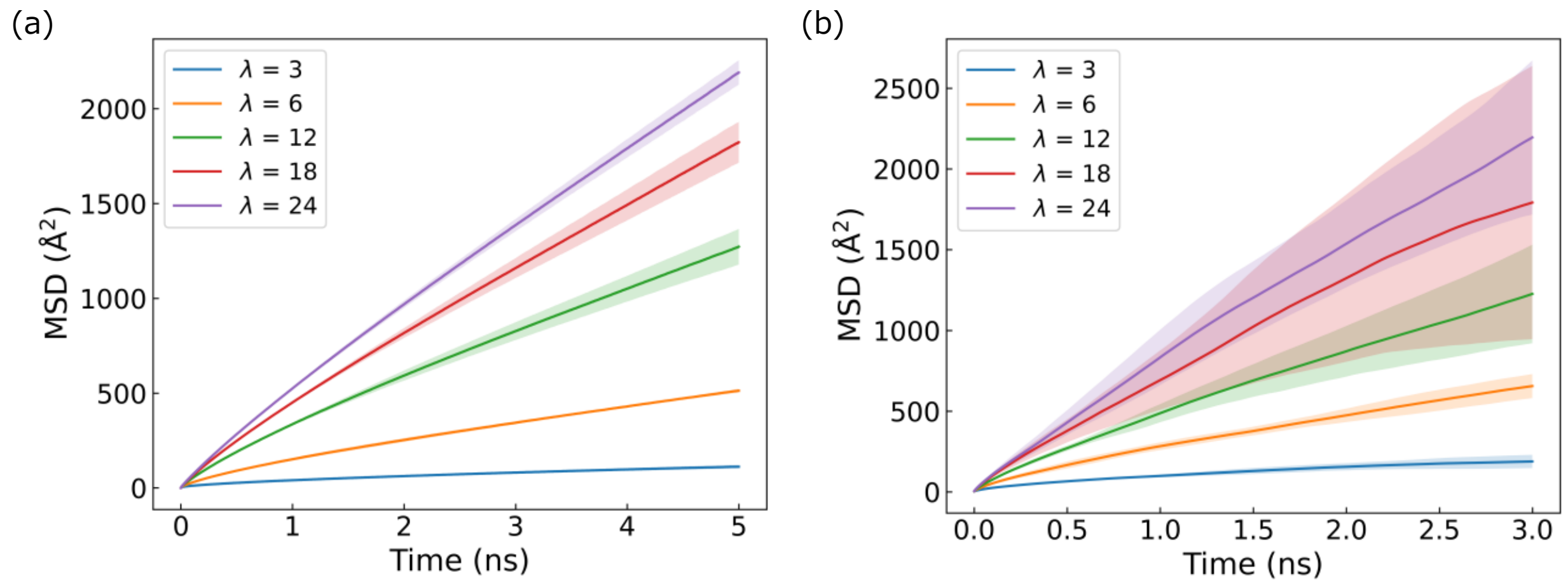

**Figure S6.** Mean-squared displacements (MSDs) of hydrogen atoms (a) and hydronium ions (b) in the hydrated Nafion systems with $\lambda$ = 3, 6, 12, 18, and 24. The shaded areas represent the standard deviations obtained from three independent simulations starting from different structures. Each MSD is calculated using the equilibrium MD trajectory from 21 ns to 31 ns.

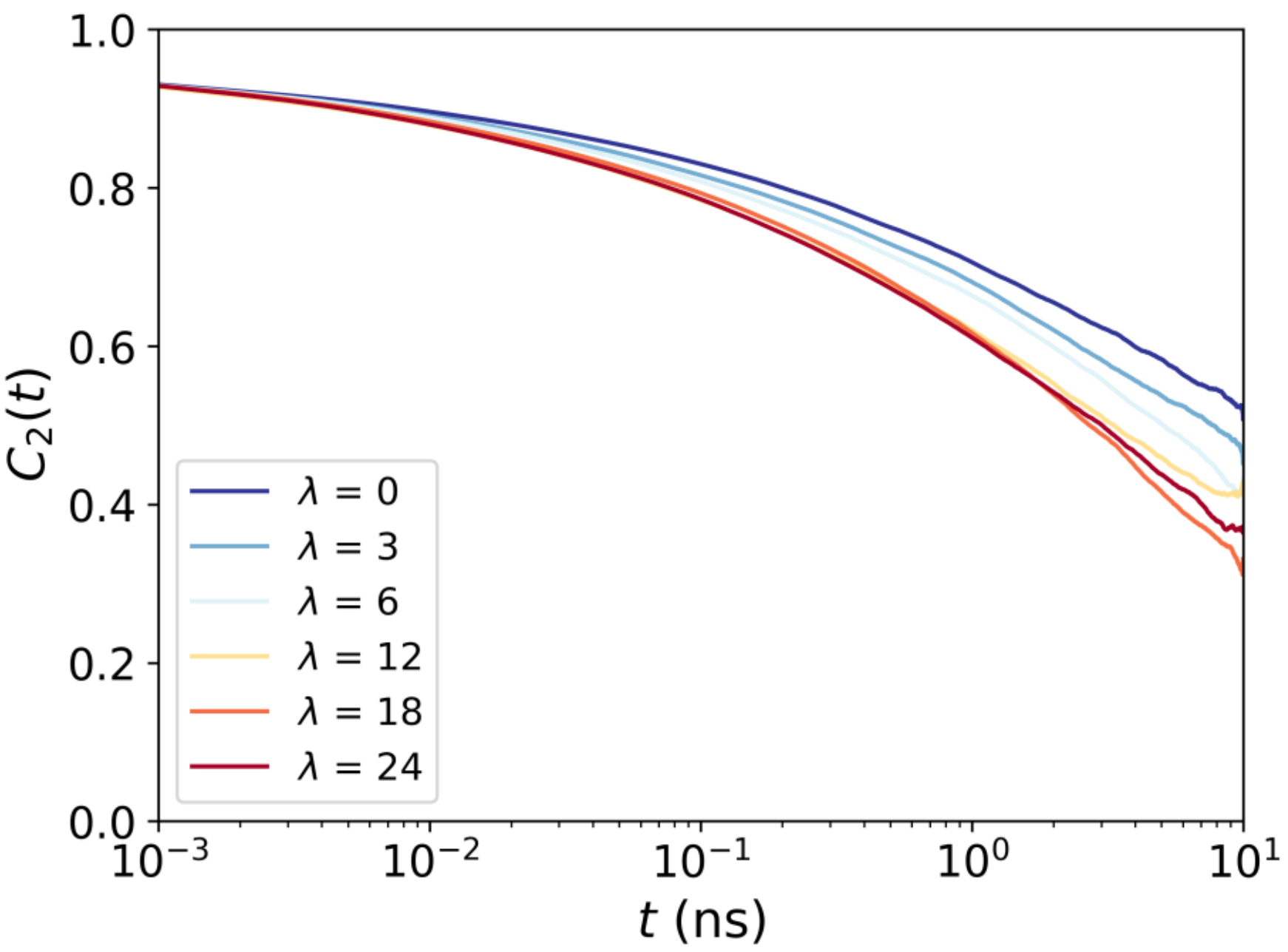

**Figure S7.** Orientational autocorrelation functions of local Nafion backbone segment vectors at different hydration levels ($\lambda$). The autocorrelation function is defined as $C_2(t) = \langle P_2(\mathbf{u}_{\text{bb}}(t) \cdot \mathbf{u}_{\text{bb}}(0))\rangle$,[14] where $\mathbf{u}_{\text{bb}}$ is a unit vector along the bond connecting two adjacent carbon atoms on the Nafion backbone and $P_2(x)$ is the second-order Legendre polynomial.

**Table S1.** Energy and Force RMSEs in Iteration 10, 13, 16, and 17[a]

| Iter. | System | Training | | Validation | |
|---|---|---|---|---|---|
| | | Energy (meV/atom) | Force (meV/Å) | Energy (meV/atom) | Force (meV/Å) |
| 10 | Nafion ($\lambda$ = 0) | 0.76 | 75.5 | 0.79 | 75.3 |
| | Nafion ($\lambda$ = 1) | 1.39 | 80.3 | 1.77 | 81.2 |
| | Nafion ($\lambda$ = 7) | 0.78 | 78.0 | 1.24 | 79.2 |
| | Nafion ($\lambda$ = 13) | 1.25 | 75.3 | 0.93 | 74.6 |
| | Water | 1.32 | 52.9 | 1.31 | 53.9 |
| 13 | Nafion ($\lambda$ = 0) | 1.09 | 79.0 | 1.42 | 79.2 |
| | Nafion ($\lambda$ = 1) | 1.67 | 81.7 | 1.66 | 82.7 |
| | Nafion ($\lambda$ = 7) | 0.97 | 79.5 | 1.58 | 80.7 |
| | Nafion ($\lambda$ = 13) | 1.26 | 77.3 | 0.94 | 76.3 |
| | Water | 1.34 | 60.7 | 1.33 | 62.2 |
| 16 | Nafion ($\lambda$ = 0) | 1.02 | 85.0 | 1.28 | 85.6 |
| | Nafion ($\lambda$ = 1) | 1.51 | 85.2 | 1.45 | 86.1 |
| | Nafion ($\lambda$ = 7) | 0.93 | 82.5 | 1.42 | 83.2 |
| | Nafion ($\lambda$ = 13) | 1.34 | 82.1 | 0.96 | 81.2 |
| | Water | 1.28 | 59.6 | 1.22 | 60.4 |
| 17 | Nafion ($\lambda$ = 0) | 0.73 | 62.5 | 0.92 | 63.3 |
| | Nafion ($\lambda$ = 1) | 1.02 | 63.5 | 1.02 | 64.6 |
| | Nafion ($\lambda$ = 7) | 0.70 | 64.0 | 1.02 | 64.8 |
| | Nafion ($\lambda$ = 13) | 0.92 | 63.6 | 0.69 | 63.2 |
| | Water | 1.05 | 47.4 | 1.01 | 48.2 |

[a]The training and validation data are split at an 8:2 ratio in each iteration. The training steps are set to 2,000,000 in iteration 17, whereas they are set to 500,000 in all other iterations.

**Table S2.** Comparison of the Numbers of Sulfo Groups, Water Molecules, and Hydronium Ions for Different Hydration Levels.

| Hydration level ($\lambda$) | # $SO_3H$ | # $H_2O$ | # $H_3O^+$ |
|---|---|---|---|
| 0 | 140 | — | — |
| 3 | 10.7 ± 2.1 | 290.7 ± 2.1 | 129.3 ± 2.1 |
| 6 | 0.64 ± 0.69 | 700.6 ± 0.7 | 139.4 ± 0.7 |
| 12 | 0.04 ± 0.20 | 1540.0 ± 0.2 | 140.0 ± 0.2 |
| 18 | 0.03 ± 0.17 | 2380.0 ± 0.2 | 140.0 ± 0.2 |
| 24 | 0.01 ± 0.11 | 3220.0 ± 0.1 | 140.0 ± 0.1 |